# Defect Prevention Approaches In Medium Scale It Enterprises

SUMA V
Asst.Professor, Dept of ISE, Dayananda Sagar College of Engineering, Bangalore, sumavgmail.com

T.R.Gopalakrishnan Nair,
Director, Research and Industry, Professor Department of Computer Science and Engineering, DSCE, Bangalore, trgnair@yahoo.com

***ABSTRACT***

*The software industry is successful, if it can draw the complete attention of the customers towards it. This is achievable if the organization can produce a high quality product. To identify a product to be of high quality, it should be free of defects, should be capable of producing expected results. It should be delivered in an estimated cost, time and be maintainable with minimum effort.*

*Defect Prevention is the most critical but often neglected component of the software quality assurance in any project. If applied at all stages of software development, it can reduce the time, cost and resources required to engineer a high quality product.*

*A small increase in the prevention measure will normally create a major decrease in total quality cost. But the main objective of quality cost analysis is not to reduce the cost, but to make sure that the cost spent are the right kind of cost and that maximizes the benefit derived from that investment. Due to quality cost analysis, the major emphasis has been shifted to prevention of defects. Also over a period of time, it is observed in most of the companies that at some optimum point, the business performance enhances, software quality increases and the cost of quality decreases due to the adoption of effective defect detection and prevention activities.*

*The scope of this paper is to provide a comprehensive view on the defect prevention techniques and practices followed in various software houses. The section 1 of the paper gives introduction to defect terminology, section 2 brings about the need for defect prevention, defect identification is briefed in section 3, section 4 tells about the classification methods followed in different companies. Section 5 describes the various practices, techniques and methodologies adopted towards defect prevention. Sections 6, 7 and 8 talks about defect measurement and analysis, benefits of defect prevention and limitations.*

## 1. INTRODUCTION

[1] A defect refers to any flaw or imperfection in a software work product or software process. The term defect refers to an error, fault or failure. [2][3]The IEEE/Standard defines the following terms as

Error: human actions that leads to incorrect result.
Fault: incorrect decision taken while understanding the given information, to solve problems or in implementation of process. A single error may lead to a single or several faults. Various errors may lead to one fault.
Failure: is inability of a function to meet the expected requirements.

With above definitions, a causal relationship among the three can be established. Thus a defect can be referred to as error or fault or failure.[4] A defect can also be defined as an issue or situation calling software change request i.e. if something is broken or not properly built or generated with a reason for not usable in certain cases, it can be defect.

Defect prevention is a process of identifying these defects, their causes and correcting them and to prevent them from recurring. [5] Test strategies can be classified into two different categories namely defect prevention technologies and defect detection technologies. DP provides the greatest cost and schedule savings over the duration of the application development efforts. [6]There are two approaches for tackling these problems and they are curative approach and preventive approach. In case of curative approach, the focus is on identifying the defects by developers and users of the software. In preventive approach, the focus is on preventing defects at the root level. DP can be applied to one or more phases of the software life cycle.

## 2. NEEDS FOR DEFECT PREVENTION

Analysis of the defects at early stages reduces the time, cost and the resources required. The knowledge of defect injecting methods and processes enable the defect prevention. Once this knowledge is practiced the quality is improved. It also enhances the total productivity.

## 3. DEFECT IDENTIFICATION

[7] There are several approaches to identify the defects like inspections, prototypes, testing and correctness proofs. Formal inspection is the most effective and expensive quality assurance technique for identifying defects at the early stages of the development. Through prototyping several requirements are clearly understood which helps in overcoming the defects. Testing is one of the least effective techniques. Those of the defects, which could have escaped by identification at the early stages, can be detected at the time of testing. Correctness proofs are also a good means of detecting especially at the coding stage. Correctness in construction is the most effective and economical method of building the software.

## 4. CLASSIFICATION OF DEFECTS

Once the defects are identified, they are classified at two different points in time namely the time at which the defect is first detected and the time when the defect has been fixed. Several models and tools are available for defect classification like ODC, which is used throughout IBM.[8] ODC essentially means that we categorize a defect into classes that collectively point to the part of the process which needs attention, much like characterizing a point in a Cartesian system of orthogonal axes by its (x,y,z) coordinates.

[9]HP (Hewlett Packard) company uses HP model which links together the defect types and origin so that it is clear which type appears to which origin. Infosys classify defects based on certain factors like logical functions, user interface, standards, and maintainability and so on. Likewise, each company has there own methodology of classifying the defects. [7] Identified defects may then fall among one of the following categories like the blocker, which prevents the engineers from testing or developing the software, the critical, which results in software crash or system hang or loss of data, the major which results in breaking down a major feature, the minor which causes a minor loss of function but can create an easy work around, the trivial, which is a cosmetic problem. Based on these categories, severity levels are assigned as urgent/show stopper, medium/work around and low/cosmetic.

## 5. DEFECT PREVENTIVE TECHNIQUES AND PRACTICES

[3] By understanding the previous definitions of defect, error, fault and failure, defects can be dealt in three categories namely

- Defects prevention through error removal
- Defect reduction through fault detection and removal
- Defect containment through failure prevention

5.1.1 Defect prevention through error removal

Defect through error sources can be removed in one or combination of following ways

Train and educate the developers.

[10] [3] about 40 to 50% of user programs contain non trivial defects. Train the people and educate them in product and domain specific knowledge. Developers should improve the development process knowledge and expertise in software development methodology as well. Introduction of disciplined personal practices like clean room approach, personal software process and team software process reduces defect rate by up to 75%.

- Use of formal methods like formal specification and formal verification.

Formal specification is concerned with producing consistent requirements specification, constrains and designs so that it reduces the chances of accidental fault injections. With formal verifications, correctness of software system is proved. Axiomatic correctness is one such method.

- Defect prevention based on tools, technologies, process and standards.

Most of the company uses object oriented methodology which supports information hiding principle and reduces interface interactions, thus reducing interface or interaction problems. Likewise by following a managed process, ensuring of appropriate process selection and conformance, enforcement of selected product and development standard also prevents defect recurrence to a large extent.

- Prevention of defects is possible by analyzing the root causes for the defects.

Root cause analysis can take up two forms namely logical analysis and statistical analysis. Logical analysis is a human intensive analysis which requires expert knowledge of product, process, development and environment. It examines logical relation between faults (effects) and errors (causes).

Statistical analysis is based on empirical studies of similar projects or locally written projects.

5.1.2. Defect reduction through fault detection and removal

Large companies go for extensive mechanisms to remove as many faults as possible under project constraints. Inspection is direct fault detection and removal technique while testing is observation of failure and fault removal. . Inspections can range from informal reviews to formal inspections. Testing phase can be subdivided as code phase of the product before the shipment and post release phase of the product. It includes all kinds of testing from unit testing to beta testing.

5.1.3. Defect containment through failure prevention

In this defect preventive approach, causal relationship between faults and resulting failures are broken and there by preventing defects, but allowing faults to reside. Techniques like recovery blocks, n-version programming, safety assurance and failure containment are used. With the use of recovery blocks, failures are detected but the underlying faults are not removed, even though the off-line activities can be carried out to identify and remove the faults in case of repeated failures. N-version programming is most applicable when timely decisions or performance is critical such as in many real time control systems. Faults in different versions are independent, which implies that it is rare to have the same fault triggered by the same input and cause the same failure among different versions. For some safety critical system, the aim is to prevent accidents where an accident is a failure with severe consequence. In addition to above said quality assurance activities, specific techniques are used based on hazards or logical preconditions for accidents like hazard elimination, hazard reduction, hazard control, damage control.

5.2. [11] Both the organization and the projects must take specific actions to prevent recurrence of defects. Some of the actions that are handled as described in Process Change Management Key Process Area are: - Goals, Commitment to perform, Ability to perform, Activities performed, Measurements and analysis and verifying implementations. The organization sets three goals like defect prevention activities which are planned, common causes of defects to seek out and to be identified, common causes of defects to be prioritized and systematically eliminated. The management owes certain commitment in order to get these goals into life. This commitment is seen as a written policy which is framed and implemented. The stipulated policy exists for the organization and for the project. It includes long term plans for funding, staffing and for the resources required for defect prevention. To improve the software processes and the products through DP activities, these results need to be reviewed and the actions are identified and addressed. For the DP to be able to perform, as per the Key Process Area, an organizational level team as well as the project level should exist. This may include teams from the Software Engineering Process Group. The software project core develops and maintains a plan for DP activities which contain the plan for task kickoffs, causal analysis meetings to be held, schedule of activities, assigned responsibilities and resources. Reviews to these are carried as per the Peer Review Key Process Area. In the kick off meetings, as per the Software Quality Management Key Process Area, the members of the team get themselves familiarized with the standards, process, procedures, methods and tools available, inputs of errors commonly introduced and recommended preventive actions for them, team assignments and software quality goals. A causal analysis meeting is a periodic review. The defects identified are analyzed to determine their root causes with the help of methods like cause/effect diagrams. The actions are proposed using techniques like Pareto analysis. The action proposal gets implemented as an action item, which is documented. The description of these data items include the person responsible for implementing it, areas affected by it, individuals who needs to be informed about its status, date when its next status is reviewed, rationale for the decisions, implementation actions, time, cost for identifying defect and correcting it and the estimated cost for not fixing it. As per Software Configuration Management Key Process Area, these data needs to be managed and controlled. The organization may have to revise its standards in process or in project's defined process according to the DP actions. On a periodic basis the team reviews, the status and the results of the organization and the project's DP activities need to be reviewed.

5.3. [10] Defects can be reduced and henceforth prevented by following certain key aspects like: - Use of prototyping approach where needs of the customer and developer becomes clearer. Preferences of emergent process against reduction list process where requirements emerge from prototyping and multiple stake holder's shared learning activities rather than requirements collected in advance. Defects can be prevented by not encouraging hasty elicitation of requirements and nominal design. Not overlooking the factors like internal cohesion, coupling and data structures, amount of change to reused code and context dependent factors, which tend to prone errors. High-risk scenarios have to be tested rigorously. Number of peer reviews, type, size and complexity of system, frequency of occurrence of defects caught has an effect on defect removal. Scenarios based reading technique consisting of union of several perspectives of inspection give a broad coverage of defects.

5.4. [12] Some company adopts quality control activities to uncover defects and have them corrected so that defect free products will be produced. Quality control in real meaning is to inspect the finished goods prior to shipment. In software applications, quality control tends to find the defects in a product by a monitoring, auditing and assessment of process. Quality control monitors and asses procedures while quality testing finds and isolate the procedure.

5.5. Defect prevention can be achieved with automation of the development process. There are several tools available right from the requirements phase to testing phase. [5] Tools available at requirements phase are quite expensive. They can be automated for consistency check but not completeness check. Tools used at this phase include requirement management tools, requirements recorders tools, requirement verifier's tools etc. the design tools include database design tools, applications design tools, visual modeling tools like Rational Rose and so on. Testing phase can be automated by the use of tools like code generation tools, code testing tools, code coverage analyzer tools. Several tools like defect tracking tools, configuration management tools and the test procedures generation tools can be used in all phases of development.

6. Defect measurement and analysis

Causal analysis and statistical defect models are the two extremes ways of measuring the status of defect preventive activities. Causal analysis is a qualitative analysis. Fish Bone diagram is used for complex cause analysis. Statistical defect modeling refereed as Reliability growth is a quantitative analysis method. It is measured in terms of number of defects remaining in the areas, failure rate of the product, short term defect detection rate etc.[8] ODC is a technique that bridges the gap between the qualitative and quantitative techniques.

## 7. BENEFITS OF DEFECT PREVENTION

[5] The existences of defect prevention strategies not only reflect a high level of test discipline maturity but also represent the most cost beneficial expenditure associated with the entire test effort. Detection of errors in the development life cycle helps to prevent the migration of errors from requirement specifications to design and from design into code. Thus test strategies can be classified into two different categories i.e. defect prevention technologies and defect detection technologies. Defect prevention provides the greatest cost and schedule savings over the duration of the application development efforts. Thus it significantly reduces the number of defects, brings down the cost for rework, makes it easier to maintain, port and reuse. It also makes the system reliable, offers reduced time and resources required for the organization to develop high quality systems. The defects can be traced back to the life cycle stage in which they were injected based on which the preventive measures are identified which in turn increases productivity. A defect preventive measure is a mechanism for propagating the knowledge of lessons learned between projects.

## 8. LIMITATIONS

[6] There is a need to develop and apply software in new and diverse domains where specific domain knowledge is lacking. In several occasions appropriate quality requirements might not be specified at first place. The conduction of inspections is labor intensive and requires high skills. Sometimes full-blown quality measurements may not have been identified at design time.

## CONCLUSIONS

Defect prevention methodologies cannot always prevent all defects from entering into the applications under test because application is very complex and it is impossible to catch all the errors. Defect detection techniques compliment defect prevention efforts and the two methodologies work hand in hand to increase the probability that the test team will meet its defined test goals and objectives. The existences of defect prevention strategies not only reflect a high level of test discipline maturity, but also represent the most cost beneficial expenditure associated with the entire test effort. Detection of errors in the development life cycle helps to prevent the migration of errors from requirement specification to design and from design into code. Defect prevention is very much vital for an organization's quality growth. The main objective of quality cost is not to reduce the cost but to invest the cost on right investment. It should not be treated as wastage of time, demanding deep involvement. Instead of, it should be considered as a saving of time, cost and the resources required. It saves a lot of rework required when the defects gets manifested at the final stages or at the post delivery period. Defect prevention should be introduced at every stage of the software life cycle to block the defects at the earliest, take corrective actions for its elimination and to avoid its reoccurrence. There are several methods, tools, techniques and practices for defect prevention but all seems to be not sufficient enough. A lot of work is still required for the defect prevention in terms of techniques to be adopted, tools to be used and policies to be written.

[This work is not a part of the software consultancy work carried out to any particular industry where the authors are involved with.]